\documentclass[12pt,journal,onecolumn]{IEEEtran}

\usepackage{epsfig,graphicx,amssymb,color}
\usepackage{amscd}
\usepackage{amsmath}
\usepackage{enumerate}
\usepackage{theorem}

\theoremstyle{plain} \theorembodyfont{\itshape}
\theoremheaderfont{\scshape}
\newtheorem{theorem}{Proposition}
\theorembodyfont{\itshape} \theoremheaderfont{\scshape}

\theoremstyle{plain} \theorembodyfont{\itshape}
\theoremheaderfont{\scshape}

\newtheorem{assumption}{Assumption}
\newtheorem{definition}{Definition}

\begin{document}

\title{Vector Precoding for Wireless MIMO Systems:\\ A Replica Analysis}

\author{
Ralf R.\ M{\"u}ller, Dongning Guo, and Aris L.\ Moustakas
\thanks{R.\ M{\"u}ller is with the Department of Electronics and Telecommunications, The Norwegian University of Science and Technology, Trondheim, Norway, e-mail:ralf@iet.ntnu.no}
\thanks{D.\ Guo is with the Department of Electrical Engineering and Computer Science,
  Northwestern University, Evanston, IL, USA, e-mail:dGuo@northwestern.edu}
\thanks{A.\ Moustakas is with the Physics Department, National and Kapodistrian University of Athens, Athens, Greece, e-mail:arislm@phys.uoa.gr}
}


\maketitle
\begin{abstract}
We apply the replica method to analyze vector precoding, a method
to reduce transmit power in antenna array communications. The
analysis applies to a very general class of channel matrices. The
statistics of the channel matrix enter the transmitted energy per
symbol via its R-transform. We find that vector precoding performs
much better for complex than for real alphabets.

As a byproduct, we find a nonlinear precoding method with
polynomial complexity that outperforms NP-hard Tomlinson-Harashima
precoding for binary modulation on complex channels if the number
of transmit antennas is slightly larger than twice the number of
receive antennas.
\end{abstract}


\newcommand{\fzv}[2]{\noindent #1 \hfill \parbox{13.2cm}{#2}}
\def\mathlette#1#2{{\mathchoice{\mbox{#1$\displaystyle #2$}}%
                               {\mbox{#1$\textstyle #2$}}%
                               {\mbox{#1$\scriptstyle #2$}}%
                               {\mbox{#1$\scriptscriptstyle #2$}}}}
\newcommand{\matr}[1]{\mathlette{\boldmath}{#1}}
\newcommand{\SINR}{{sig\-nal--to--dis\-tor\-tion ratio}}
\newcommand{\SNR}{{sig\-nal--to--noise ratio}}
\renewcommand{\j}{{\rm j}}
\newcommand{\RR}{\mathbb{R}}
\newcommand{\CC}{\mathbb{C}}
\newcommand{\NN}{\mathbb{N}}
\newcommand{\ZZ}{\mathbb{Z}}
\newcommand{\pic}{\pi}
\newcommand{\deltaf}{\delta}
\def\argmin{\mathop{\rm argmin}}
\def\argmax{\mathop{\rm argmax}}
\newcommand{\diag}{{\rm diag}}
\def\expect{\mathop{\mbox{\large ${\rm E}$}}}
\newcommand{\Hilbert}{{\cal H}}
\newcommand{\Landau}[1]{{\cal O}\left(#1\right)}
\newcommand{\landau}[1]{o\left(#1\right)}
\newcommand{\Var}[1]{{\rm Var}\left\{#1\right\}}
\newcommand{\comp}[1]{{#1^{\rm c}}}
\newcommand{\hermite}[1]{{#1^{\rm H}}}
\newcommand{\transp}[1]{{#1^{\rm T}}}
\newcommand{\conj}[1]{{#1^{\ast}}}
\newcommand{\e}{{\rm e}}
\newcommand{\iu}{{\rm j}}
\newcommand{\vnull}{{\rm \bf 0}}
\newcommand{\I}{{\rm \bf I}}
\newcommand{\chol}[1]{{\rm chol}\left(#1\right)}
\newcommand{\h}{{\rm ent}}
\newcommand{\prob}[2]{{\rm p}_{#1}\!\!\left( #2 \right) }
\newcommand{\Prob}[2]{{\rm P}_{#1}\!\!\left( #2 \right) }
\newcommand{\proba}[2]{\breve{\rm p}_{#1}\!\!\left( #2 \right) }
\newcommand{\Proba}[2]{\breve{\rm P}_{#1}\!\!\left( #2 \right) }
\newcommand{\Probi}[2]{{\rm P}_{#1}^{-1}\!\left( #2 \right) }
\newcommand{\StT}[2]{{\rm G}_{#1}\!\left( #2 \right) }
\newcommand{\StTi}[2]{{\rm G}_{#1}^{-1}\!\left( #2 \right) }
\newcommand{\ST}[2]{{\rm S}_{#1}\!\left( #2 \right) }
\newcommand{\RT}[2]{{\rm R}_{#1}\!\left( #2 \right) }
\newcommand{\YT}[2]{{\Upsilont}_{#1}\!\left( #2 \right) }
\newcommand{\YTi}[2]{{\Upsilont}_{#1}^{-1}\left( #2 \right) }
\newcommand{\Q}{{\rm Q}}
\newcommand{\load}{{\beta}}
\newcommand{\quant}{{\rm quant}}
\newcommand{\sign}{{\rm sign}}
\newcommand{\dirac}[1]{\deltaf \! \left( #1 \right)}
\newcommand{\kron}[1]{\deltaf \! \left[ #1 \right]}
\newcommand{\MomGen}[2]{{\Phif}_{#1}\left( #2 \right) }
\newcommand{\tsf}{\Theta}
\newcommand{\nmtsf}{{\theta_{\min}}}
\newcommand{\mlatt}{\mu_{\ln \! \Att}}
\newcommand{\slatt}{\sigma_{\ln \! \Att}}
\newcommand{\varnoise}{\sigma_\noise^2}
\newcommand{\aux}{\Upsilon}
\newcommand{\define}{\stackrel{\triangle}{=}}
\newcommand{\D}{\displaystyle}
\newcommand{\eq}[1]{(\ref{#1})}
\newcommand{\eqs}[2]{(\ref{#1}) and (\ref{#2})}
\newcommand{\eqd}[3]{(\ref{#1}), (\ref{#2}), and (\ref{#3})}
\newcommand{\eqv}[4]{(\ref{#1}), (\ref{#2}), (\ref{#3}), and (\ref{#4})}
\newcommand{\azeta}{\left(1-\sqrt{\zeta}\right)^2}
\newcommand{\bzeta}{\left(1+\sqrt{\zeta}\right)^2}
\newcommand{\bin}[2]{{\left(\!\begin{array}{c} #1 \\ #2 \end{array}\!\right)}}
\newcommand{\xyplot}[3]{\begin{tabular}{c}\begin{turn}{90}\begin{tabular}{c}
    #3 \\[-2mm] \begin{turn}{-90} #1 \end{turn}\end{tabular}\end{turn}
    \\[-2mm] \hphantom{\begin{turn}{90} #3 \end{turn}{90}}
    \hspace*{\parskip} #2 \end{tabular}}
\section{Introduction}

Wireless multiple-input multiple-output (MIMO) systems offer the
possibility to increase data rate over conventional wireless
communications without need for more physical radio spectrum by
means of multiple antenna elements at both transmitter and
receiver side. Since the pioneering work in the field
\cite{foschini:98,telatar:99}, countless implementations for those
MIMO systems have been proposed. They can be classified by the
side where the signal processing takes place. Depending on the
proposed system solution, there can be need for major signal
processing at the receiver side, the transmitter side or both of
them. This work is concerned with systems where sophisticated
signal processing is required solely at the transmitter side. This
is advantageous for transmitting data to low-cost or
battery-driven devices such as cell-phones and PDAs.

It is an unavoidable feature of wireless MIMO systems that signals
sent at different antenna elements of the transmit array are
received with severe crosstalk at the respective antenna elements
of the receive array. In order to compensate for this crosstalk,
one can use linear joint transmitter processing, also known as
linear vector precoding, as suggested in \cite{vojcic:98,peel:05}.
This comes, however, at the expense of the need for an increased
transmit power in order to maintain the distance properties of the
signal constellation. A more sophisticated method for transmitter
processing is nonlinear vector precoding, in this work simply
referred to as vector precoding. It is based on the concept of
Tomlinson-Harashima preecoding \cite{tomlinson:71,harashima:72}
which was originally proposed to combat intersymbol interference.
It was proposed for use in context of MIMO systems in
\cite{windpassinger:04,hochwald:05}. For a general survey on
vector precoding the reader is referred to \cite{fischer:02}.

In this work, we are mainly concerned with the performance
analysis of vector precoding. To the best of our knowledge, there
is no published literature on the performance analysis of
nonlinear vector precoding by analytical means. This paper aims to
pave the way a first step forward towards this direction employing
the replica method which was originally invented for the analysis
of spin glasses in statistical physics \cite{mezard:87,fischer:91}
and has become increasingly powerful to address problems in
wireless communications and coding theory \cite{nishimori:01}. We
use the analytical results developed in this paper to compare
real-valued vector precoding with complex-valued vector precoding
as well as with some hybrid forms of   it which are newly proposed
in this work.

The paper is composed of five more sections. Section~\ref{VP}
introduces vector precoding from a general point of view. This
point of view is more general than the way vector precoding is
dealt with in the references mentioned earlier, but it is well
suited to the replica analysis to follow. Section~\ref{PS}
formulates vector precoding as a non-convex quadratic programming
problem and introduces the technical assumptions that we require
for the analytical analysis. Section~\ref{GR} derives the general
replica symmetric solution to any non-convex quadratic programming
problem for which the search space can be factorized into
Kronecker products of scalar sets in the limit of a large number
of dimensions of the search space. Section~\ref{PR} specializes
the general results to MIMO channels with channel matrices
composed of independent identically distributed entries and
various choices for the relaxation of the symbol alphabet.
Section~\ref{C} summarizes the main conclusions. Particularly
technical derivations are placed in the two appendices.

\section{Vector Precoding}
\label{VP}

Vector precoding aims to minimize the transmitted power that is
associated with the transmission of a certain data vector $\matr
s\in{\cal S}^K$ of length $K$. For that purpose, the original
symbol alphabet ${\cal S}$ is relaxed into the alphabet ${\cal
B}$. The data representation in the relaxed alphabet is redundant.
That means that several symbols in the relaxed alphabet represent
the same data. Due to the redundant representation, we can now
choose that representation of our data which requires the least
power to be transmitted. This way of saving transmit power is what
we call vector precoding.

\begin{figure}[p]
\centerline{\epsfig{file=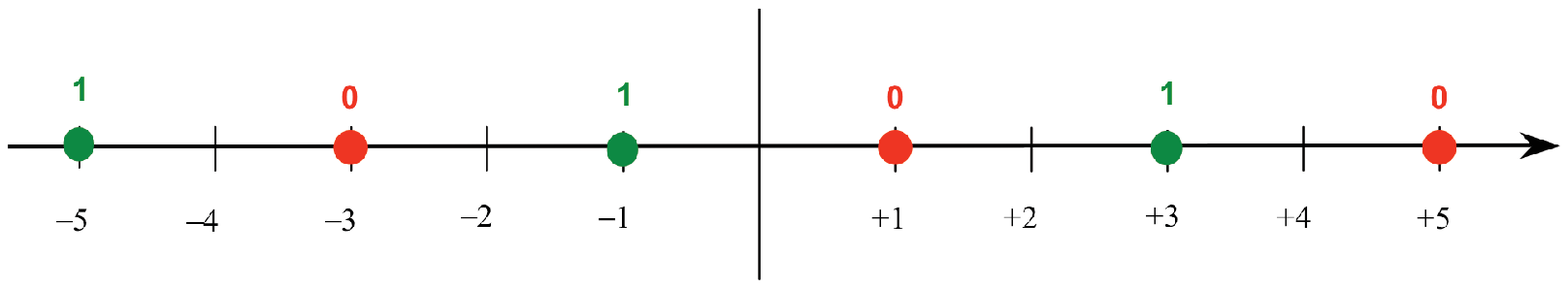,width=\columnwidth}}
\caption{\label{1dlattice} 2 one-dimensional equally spaced
integer lattices representing the two binary states 0 and 1,
respectively.}
\end{figure}

That means, for any $s \in {\cal S}$ there is a set ${\cal B}_{s}
\subset {\cal B}$ such that all elements of ${\cal B}_{s}$
represent the data $s$. Take binary transmission as an example,
i.e.\ ${\cal S}=\{1,0\}$. Without vector precoding, it is most
common to choose ${\cal B}_0=\{+1\}$ and ${\cal B}_1=\{-1\}$. This
modulation is called binary phase shift keying. For binary
modulation, vector precoding is the idea to have  ${\cal B}_0
\supset \{+1\}$ and ${\cal B}_1\supset \{-1\}$, i.e.\ to allow for
supersets of the binary constellation. A popular choice for those
supersets is due to Tomlinson and Harashima
\cite{tomlinson:71,harashima:72}, see also Fig.~\ref{1dlattice}.
Here, we have ${\cal B}_0 =4\ZZ+1$ and ${\cal B}_1=4\ZZ-1$.

In order to avoid ambiguities, we should have
\begin{equation}
{\cal B}_{i} \cap {\cal B}_{j} = \emptyset \qquad \forall i\ne j.
\end{equation}
In addition, one would like to design the sets ${\cal B}_{i}$ such
that the distance properties between the presented information are
preserved. This is easily achieved by letting the sets ${\cal
B}_{i}$ to be distinct sub-lattices of ${\cal B}$. However, we are
not concerned with these design issues here. We aim to analyze the
power saving achieved by a particular choice of the sets ${\cal
B}_{i}$. This goal is achieved using the replica method invented
in statistical physics.

\section{Problem Statement}
\label{PS}

Let $\matr s=[s_1,\dots,s_K]^{\rm T}$ denote the information to be
encoded. Let $\matr t=\matr {Tx}$ be the vector being sent. Then,
the precoding problem can be written as the minimization of the
following quadratic form
\begin{equation}
\label{qp} \min\limits_{\matr x\in {\cal X}} ||\matr {Tx}||^2 =
\min\limits_{\matr x\in {\cal X}} \matr x^{\dagger}\matr{Jx}
\end{equation}
over the discrete set
\begin{equation}
{\cal X} = {\cal B}_{s_1} \times  {\cal B}_{s_2} \times \cdots
\times  {\cal B}_{s_K}
\end{equation}
with $\matr J=\matr {T^\dagger\matr T}$. This type of problem is
known in context of optimization as non-convex quadratic
programming.

In order to allow for analytical tractability, we need the
follwing definition and assumptions:
\begin{definition}[R-transform]
Let $\Prob{}x$ denote an arbitrary probability distribution. Let
\begin{equation}
\label{defST} m(s)=\int\frac{{\rm d}\Prob{}x}{x-s}.
\end{equation}
Then, the R-transform of  $\Prob{}x$ is
\begin{equation}
R(w)=m^{-1}(-w)-\frac1w
\end{equation}
with $m^{-1}(w)$ denoting the inverse of $m(s)$ with respect to
composition, i.e.\ $m^{-1}(m(s))=s$.
\end{definition}
\begin{assumption}[self-averaging property]
We have
\begin{equation}
\lim\limits_{K\to\infty}\Pr\left(\frac1K\left| \min\limits_{\matr
x\in {\cal X}} \matr x^{\dagger}\matr{Jx} -  \expect\limits_{\matr
J} \min\limits_{\matr x\in {\cal X}} \matr
x^{\dagger}\matr{Jx}\right|>\epsilon\right)=0
\end{equation}
for all $\epsilon>0$, i.e.\ convergence in probability.
\end{assumption}

\begin{assumption}[replica continuity]
For all $\beta>0$, the continuation of the function
\begin{equation}
f(n)={\prod\limits_{a=1}^n \sum\limits_{\matr x_a\in {\cal X}}
{\rm e}^{\,-\beta\matr x^{\dagger}_a\matr{Jx}_a} }
\end{equation}
onto the positive real line is equal to
\[
\left( \sum\limits_{\matr x\in {\cal X}} {\rm e}^{\,-\beta\matr
x^{\dagger}\matr{Jx}} \right)^n
\]
in the right-sided vicinity of $n=0$.
\end{assumption}

\begin{assumption}[unitary invariance]
The random matrix $\matr J$, can be decomposed into
\begin{equation}
\matr J=\matr{ODO}^\dagger
\end{equation}
such that the matrices $\matr D$ and $\matr O$ are diagonal and
Haar distributed, respectively. Moreover, as $K\to\infty$, the
asymptotic eigenvalue distribution of $\matr J$ converges to a
non-random distribution function which can be uniquely
characterized by its R-transform $R(w)$.
\end{assumption}

\begin{assumption}[replica symmetry]
When applying the replica method to solve the saddle-point
equations, we will assume that the extremal point is invariant to
permutations of the replica index. For a detailed discussion of
replica symmetry, the reader is referred to the literature of spin
glasses, e.g.\ \cite{mezard:87,fischer:91}.
\end{assumption}

The first three assumptions are rather technical and should hold
well in the application we are addressing. The validity of replica
symmetry for the minimum value of $\matr x^\dagger\matr{Jx}$ is an
approximation, which is made for analytical tractability and, for
sake of simplicity, no further justification is made here. It
should be pointed out though that, even when replica symmetry is
not valid the correct value of quantities such as $E_{\rm s}$ do
not differ much from the corresponding values evaluated within the
replica-symmetric assumption, cf.\
\cite{kirkpatrick:78,marinari:94}.
\section{General Result}
\label{GR}

In this section, we derive a general solution to the non-convex
quadratic programming problem \eq{qp} in the limit of a large
number of dimensions $K$. We find the following result:
\begin{theorem}
Let Assumptions 1 to 4 hold. Let $\Prob{s\,}s$ denote the limit of
the empirical distribution function of the information symbols
$s_1,\dots,s_K$ as $K\to\infty$. Moreover, let the parameters $q$
and $b$ be solutions to the following pair of coupled fixed-point
equations
\begin{eqnarray}
\label{theoq1}
q&=&\int\!\!\!{\int \left|\argmin_{x\in {\cal B}_{s}}\left|z\sqrt{\frac{qR^\prime(-b)}{2R^2(-b)}}-x\right|\right|^2{\rm D}z}{\rm d}\Prob {\!s\,}{s}\\
\label{theob1} b&=& \int\!\!\!{\int \Re\left\{\argmin_{x\in {\cal
B}_{s}}\left|z\sqrt{\frac{qR^\prime(-b)}{2R^2(-b)}}-x\right|z^\ast\right\}\frac{{\rm
D}z{\rm d}\Prob {\!s\,}{s}} {\sqrt{2qR^\prime(-b)}}}.
\end{eqnarray}
with ${\rm D}z=\exp(-z^2/2)/(2\pi){\rm d}z$ being the complex
Gaussian measure. Then, if $0<b<\infty$, we have
\begin{equation}
\label{theoEs1} \frac1K\min\limits_{\matr x\in{\cal X}} \matr
x^\dagger\matr{Jx} \to q\, \frac{{\partial}}{{\partial}b}\, b
R(-b)
\end{equation}
in probability as $K\to\infty$.
\end{theorem}

The remainder of this section is dedicated to the derivation of
Proposition 1. Further sections will not make reference to the
remainder of Section~\ref{GR}.

With Assumptions 1 and 2, we find for the average transmitted
energy per symbol in the large system limit
\begin{eqnarray}
E_{\rm s}\!\!\! &=& \!\!\lim\limits_{K\to\infty} \frac1K\min\limits_{\matr x\in {\cal X}} \matr x^{\dagger}\matr{Jx}\\
&=& \!\!-\lim\limits_{K\to\infty}\lim\limits_{\beta\to\infty}\frac1{\beta K}\expect\limits_{\matr J} \log\sum\limits_{\matr x\in {\cal X}} {\rm e}^{\,-\beta\matr x^{\dagger}\matr{Jx}}\\
&=&\!\!-\lim\limits_{K\to\infty}\lim\limits_{\beta\to\infty}\frac1{\beta K}\lim\limits_{n\to0}\frac\partial{\partial n}\log \expect\limits_{\matr J} \left( \sum\limits_{\matr x\in {\cal X}}{\rm e}^{\,-\beta\matr x^{\dagger}\matr{Jx}}\right)^n \\
&=&\!\! -\lim\limits_{\beta\to\infty}\frac1{\beta
}\lim\limits_{n\to0}\frac\partial{\partial n}
\underbrace{\lim\limits_{K\to\infty}\frac1K\log\expect\limits_{\matr
J}\prod\limits_{a=1}^n \sum\limits_{\matr x_a\in {\cal X}} {\rm
e}^{\,-\beta\matr x^{\dagger}_a\matr{Jx}_a} }_{\define
\Xi_n}.\label{eq518} \nonumber
\end{eqnarray}
where the argument of the logarithm in \eq{eq518} is given
by\footnote{The notation $\sum_{\{x_a\}}$ is used as shortcut for
$\sum_{x_1}\sum_{x_2}\cdots \sum_{x_n}$.}
\begin{eqnarray}
\Xi_n&=& \lim\limits_{K\to\infty}\frac1K\log\expect\limits_{\matr
J} \sum\limits_{\{\matr x_a\in {\cal X}\}}
\exp\left[-\beta \sum\limits_{a=1}^n \matr x_a^{\dagger}\matr{Jx}_a \right]\\
&=& \lim\limits_{K\to\infty}\frac1K\log\expect\limits_{\matr J}
\sum\limits_{\{\matr x_a\in {\cal X}\}} \exp\left[{\rm tr}
\left(-\beta \matr J \sum\limits_{a=1}^n \matr x_a \matr
x_a^{\dagger}\right)\right]. \nonumber
\end{eqnarray}

Using Assumption 3, we can integrate over the Haar distributed
eigenvectors of $\matr J$. This integration was studied by
Harish-Chandra \cite{harish:57} and Itzykson \& Zuber
\cite{itzykson:80} in the mathematics and physics literature,
respectively. It was applied in context of wireless communications
in \cite{takeda:06}. Guionnet and Ma\"ida \cite{guionnet:05} solve
this integral in terms of the R-transform $R(w)$ of the asymptotic
eigenvalue distribution of $\matr J$. Following their approach
yields
\begin{equation}
\Xi_n= \lim\limits_{K\to\infty}\frac1K\log\sum\limits_{\{\matr
x_a\in {\cal X}\}} \exp\left[-K\sum\limits_{a=1}^n
\int\limits_0^{\lambda_a} R(- w)dw \right] \label{intX}
\end{equation}
with $\lambda_i$ denoting the $n$ positive eigenvalues of
\begin{equation}
\beta\sum\limits_{a=1}^n \matr x_a \matr x_a^{\dagger}.
\end{equation}

\newcommand{\skp}{^\dagger}
The eigenvalues $\lambda_i$ are completely determined by the inner
products
\begin{equation}
\label{defQR} K Q_{ab} = \matr x_a \skp \matr x_b \define
\sum\limits_{k=1}^K x_{ak}^\ast x_{bk}.
\end{equation}

In order to perform the summation in \eq{intX}, the
$Kn$-di\-men\-sio\-nal space spanned by the replicas is split into
subshells
\begin{equation}
S\{Q\} \define \left\{\matr x_1,\dots,\matr x_n\left|\matr
x_a\skp\matr x_b=KQ_{ab}\right.\right\}
\end{equation}
where the inner product of two different replicated vectors $\matr
x_a$ and $\matr x_b$ is constant in each subshell.\footnote{The
notation $f\{Q\}$ expresses dependency of the function $f(\cdot)$
on $Q_{ab}\, \forall a,b$.} With this splitting of the space, we
find\footnote{The notation$\prod_{a,b}$ is used as shortcut for
$\prod_{a=1}^n\prod_{b=1}^n$.}
\begin{equation}
\label{domsh} \Xi_n=
\lim\limits_{K\to\infty}\frac1K\log\int\limits_{\CC^{n^2}}{\rm
e}^{K{\cal I}\{Q\}}{\rm e}^{- K{\cal
G}\left\{Q\right\}}\prod\limits_{a,b}{\rm d}Q_{ab},
\end{equation}
where
\begin{eqnarray}
{\cal G}\{Q\} = \sum\limits_{a=1}^n \int\limits_0^{\lambda_a\{Q\}}
R(-w)\,{\rm d} w
\end{eqnarray}
and
\begin{equation}
{\rm e}^{K{\cal I}\{Q\}} = \sum\limits_{\{\matr x_a\in\cal X\}}
\prod\limits_{a,b} \deltaf\left( \matr x_a\skp \matr x_b -
KQ_{ab}\right)
\end{equation}
denotes the probability weight of the subshell composed of
two-dimensional Dirac-functions in the complex plane. This
procedure is a change of integration variables in multiple
dimensions where the integration of an exponential function over
the replicas has been replaced by integration over the variables
$\{Q\}$. In the following the two exponential terms in \eq{domsh}
are evaluated separately.

First, we turn to the evaluation of the measure ${\rm e}^{K{\cal
I}\left\{Q\right\}}$. The Fourier expansion of the Dirac measure
\begin{equation}
\delta\left({\matr x_a\skp\matr x_b} - KQ_{ab}\right) =
\int\limits_{\cal J}\exp\left[\tilde{Q}_{ab}\left( {\matr
x_a\skp\matr x_b}- KQ_{ab}\right)\right] \frac{{\rm
d}\tilde{Q}_{ab}}{2\pic {\rm j}}
\end{equation}
with ${\cal J}=(t-{\rm j}\infty;t+{\rm j}\infty)$, gives
\begin{eqnarray}
{\rm e}^{K{\cal I}\left\{Q\right\}} \!\!\!&=&\!\!\!
\sum\limits_{\{\matr x_a\in\cal X\}}  \prod\limits_{a, b} \int\limits_{\cal J}{\rm e}^{\tilde{Q}_{ab}\left({\matr x_a\skp\matr x_b}-K Q_{ab}\right)}\frac{{\rm d}\tilde{Q}_{ab}}{2\pic {\rm j}}\\
&=& \!\!\! \int\limits_{{\cal J}^{n^2}} {\rm e}^{\log
\prod\limits_{k=1}^KM_k\left\{\tilde{Q}\right\}-K\sum\limits_{a,
b}\tilde{Q}_{ab}Q_{ab}}  \prod\limits_{a, b} \frac{{\rm
d}\tilde{Q}_{ab}}{2\pic {\rm j}} \label{saddlemu}
\end{eqnarray}
with
\begin{equation}
M_k\left\{\tilde{Q}\right\} = \sum\limits_{\{x_a\in{\cal
B}_{s_k}\}}   {\rm e}^{ \sum\limits_{a,b}\tilde{Q}_{ab} x_{a}^\ast
x_{b}}.
\end{equation}
In the limit of $K\to\infty$ one of the exponential terms in
\eq{domsh} will dominate over all others. Thus, only the maximum
value of the correlation $Q_{ab}$ is relevant for calculation of
the integral.

At this point, we assume replica symmetry. This means, that in
order to find the maximum of the objective function, we consider
only a subset of the potential possibilities that the variables
$Q_{ab}$ could take. In particular, we are interested in the most
general form of the positive semidefinite matrix $\matr Q$ with
permutational symmetry when exchanging its replica indices.
Therefore, we need a matrix with all off-diagonal elements equal
to each other. Thus, we restrict them to the following two
different possibilities
 $Q_{ab}=q,\forall a\ne b$ and $Q_{aa} = q+b/\beta,\forall a$ where $b\ge 0$ since $\matr Q$ has to be positive semidefinite.
One case distinction has been made to distinguish correlations
$Q_{ab}$ which correspond to correlations between different and
identical replica indices, respectively. We apply the same idea to
the correlation variables in the dual (Fourier) domain and set
with a modest amount of foresight
$\tilde{Q}_{ab}=\beta^2f^2/2,\forall a\ne b$ and
$\tilde{Q}_{aa}=\beta^2f^2/2-\beta e,\forall a$. Note that despite
the fact that $\matr Q$ is complex, in general, its values at the
saddle-point are in fact real.

At this point the crucial benefit of the replica method becomes
obvious. Assuming replica continuity, we have managed to reduce
the evaluation of a continuous function to sampling it at integer
points. Assuming replica symmetry we have reduced the task of
evaluating infinitely many integer points to calculating four
different correlations (two in the original and two in the Fourier
domain).

The assumption of replica symmetry leads to
\begin{equation}
\sum\limits_{a, b} \tilde{Q}_{ab}Q_{ab} = \frac{n(n-1)}2\,
\beta^2f^2q + n \left(\frac{\beta f^2}2-e\right)\left(\beta
q+b\right) \label{lineq}
\end{equation}
and
\begin{equation}
M_k(e,f) = \sum\limits_{\{x_a\in{\cal B}_{s_k}\}}  {\rm
e}^{\frac\beta2\sum\limits_{a=1}^n(\beta
f^2-2e)|x_{a}|^2+2\sum\limits_{b=a+1}^n \beta
f^2\Re\left\{x_{a}^\ast x_{b}\right\}} \label{lastMk}
\end{equation}
Note that the prior distribution enters the free energy only via
\eq{lastMk}. We will focus on this later on after having finished
with the other terms.

For the evaluation of ${{\cal G}\{Q\}}$ in \eq{domsh}, we can use
the replica symmetry to explicitly calculate the eigenvalues
$\lambda_i$. Considerations of linear algebra lead to the
conclusion that the eigenvalues $b$ and $b+\beta nq$ occur with
multiplicities $n-1$ and 1, respectively. Thus we get
\begin{eqnarray}
{\cal G}(q,b)= (n-1) \int\limits_0^{b} R(-w)\,{\rm d} w +
\int\limits_0^{b+\beta nq} R(-w)\,{\rm d} w. \label{ext1}
\end{eqnarray}
Since the integral in \eq{domsh} is dominated by the maximum
argument of the exponential function, the derivatives of
\begin{equation}
{\cal G}\{Q\} + \sum\limits_{a, b} \tilde{Q}_{ab}Q_{ab}
\label{tedr}
\end{equation}
with respect to $q$ and $b$ must vanish as $K\to\infty$.\footnote{
It turns out that when $\lim_{n\to0} \partial_n \Xi_n$ is
expressed in terms of $e,f,q,b$, the relevant extremum is in fact
a maximum and not a minimum. This is due to the fact that when
drops below unity, the minima of a function become maxima and
vice-versa. For a detailed analysis of this technicality, see
\cite{parisi:80}. } Taking derivatives after plugging \eq{lineq}
and \eq{ext1} into \eq{tedr}, gives
\begin{eqnarray*}
\beta nR(-b-\beta nq) + \frac{n(n-1)}2\,\beta^2f^2 + \beta n \left(\frac{\beta f^2}2-e\right) &=& 0\\
(n-1)R(-b)+R(-b-\beta nq) + n \left(\frac{\beta f^2}2-e\right) &=&
0
\end{eqnarray*}
solving for $e$ and $f$ gives
\begin{eqnarray}
\label{resG}
e&=& R(-b)\\
f&=& \sqrt{2\frac{R(-b)-R(-b-\beta nq)}{\beta n}}
\end{eqnarray}
with the limits for $n\to 0$
\begin{eqnarray}
\label{resF}
f &\stackrel{n\to0}{\longrightarrow}&\sqrt{2q R^\prime(-b)}\\
n\,\frac{{\partial}f}{{\partial }
n}&\stackrel{n\to0}{\longrightarrow}& 0 .
\end{eqnarray}

Consider now the integration over the prior distribution in the
moment-generating function. Consider \eq{lastMk} giving the only
term that involves the prior distribution and apply the complex
Hubbard-Stratonovich transform
\begin{equation}
{\rm e}^{\frac{|x|^2}2} = \frac1{{2\pi}}\int\limits_{\CC}{\rm
e}^{\Re\{xz^\ast\} -\frac {|z|^2}2} {\rm d}z =  \int{\rm
e}^{\Re\{xz\}} {\rm D}z.
\end{equation}
Then, we find with \eq{lastMk}
\begin{eqnarray}
M_k(e,f) &=& \label{anoteqc}
\sum\limits_{\{x_a\in {\cal B}_{s_k}\}}  {\rm e}^{\frac{\beta^2f^2}2\left|\sum\limits_{a=1}^n x_{a}\right|^2-\sum\limits_{a=1}^n\beta e\left|x_{a}\right|^2}\\
&=& \sum\limits_{\{x_a\in{\cal B}_{s_k}\}}  \int {\rm e}^{\beta\sum\limits_{a=1}^nf\Re\left\{x_{a}z^\ast\right\}- e\left|x_{a}\right|^2}{\rm D}z\\
&=& \int\!\left(\sum\limits_{x\in{\cal B}_{s_k}} {\rm e}^{\beta
f\Re\{xz^\ast\}- \beta e|x|^2}\right)^n{\rm D}z
\end{eqnarray}
Moreover, for $K\to\infty$, we have by the law of large numbers
\begin{eqnarray}
\log M(e,f) &= & \frac 1K \log \prod\limits_{k=1}^K M_k(e,f)\\
\nonumber &&\hspace*{-20mm}\to\int \log
\int\!\left(\sum\limits_{x\in{\cal B}_{s}} {\rm e}^{\beta
f\Re\{z^\ast x\}-\beta e |x|^2}\right)^n{\rm D}z {\rm
d}\Prob{\!s\,}{s}.
\end{eqnarray}

In the large system limit, the integral in \eq{saddlemu} is
dominated by that value of the integration variable which
maximizes the argument of the exponential function. Thus, partial
derivatives of
\begin{equation}
\log M(e,f) - \frac{n(n-1)}2\, f^2\beta^2q - n \left(\frac{\beta
f^2}2-e\right)(b+\beta q)
\end{equation}
 with respect to $f$ and $e$ must vanish as $K\to\infty$.

An explicit calculation of the two derivatives gives the following
expressions for the macroscopic parameters $q$ and $b$
\begin{eqnarray}
\label{theob} b&=&
\frac{1}{\sqrt{2qR^\prime(-b)}}\int\!\!\!{\int\frac{\sum_{x\in{\cal
B}_s} \Re\{z^\ast x\}{\rm
e}^{\beta\sqrt{2qR^\prime(-b)}\Re\{z^\ast x\}-\beta
R(-b)|x|^2}}{\sum_{x\in{\cal B}_s}{\rm
e}^{\beta\sqrt{2qR^\prime(-b)}\Re\{z^\ast x\}-\beta R(-b)|x|^2
}}{\rm D}z}\,{\rm d}\Prob {\!s\,}{s}
\\
\label{theoq} q &=&\int\!\!\!{\int\frac{\sum_{x\in{\cal B}_s}
|x|^2{\rm e}^{\beta\sqrt{2qR^\prime(-b)}\Re\{z^\ast x\}-\beta
R(-b)|x|^2}}{\sum_{x\in{\cal B}_s} {\rm
e}^{\beta\sqrt{2qR^\prime(-b)}\Re\{z^\ast x\}-\beta
R(-b)|x|^2}}\,{\rm D}z}\,{\rm d}\Prob {\!s\,}{s} -\frac b\beta.
\end{eqnarray}
Moreover, we find
\begin{eqnarray}
\label{eqb} \lim\limits_{n\to 0} \frac{{\partial}b}{{\partial}n}
&=& 0.
\end{eqnarray}
Finally, the fixed point equations \eqs{theoq}{theob} simplify via
the saddle point integration rule to \eqs{theoq1}{theob1}. Note
that the minimization with respect to the symbol $x$ splits the
integration space of $z$ into the Voronoi regions defined by the
(appropriately scaled) signal constellation ${\cal B}_{s}$.

Returning to our initial goal, the evaluation of the average
transmitted energy per symbol, and collecting our previous
results, we find
\begin{eqnarray}
E_{\rm s}\!\!\! &=&\!\!\! -\lim\limits_{\beta\to\infty}\frac1{\beta } \lim\limits_{n\to0}\frac\partial{\partial n} \,\Xi_n\\
&=&\!\!\! \lim\limits_{\beta\to\infty}
\frac1\beta\lim\limits_{n\to0}\frac\partial{\partial n} (n-1)
\int\limits_0^{b} R(-w)\,{\rm d} w + \!\!\int\limits_0^{b+\beta
nq}\!\!\! R(-w)\,{\rm d} w  \nonumber
\\ \nonumber && \!\!
- \log M(e,f) +\textstyle \frac{n(n-1)}{2} f^2\beta^2q +\frac n{2} (f^2\beta-2e)(b+\beta q)\\
&=& \!\!\! \lim\limits_{\beta\to\infty}
\frac1\beta{\int\limits_0^{b} R(-w)\,{\rm d} w - \frac
b\beta\,R(-b)+qb R^\prime(-b)}
-\frac1\beta{\int\!\!\!\int\log\sum\limits_{x\in{\cal B}_{s}} {\rm
e}^{\beta f\Re\{z^\ast x\}-\beta e |x|^2}{\rm D}z\,{\rm
d}\Prob{\!s\,}{s}}.
\end{eqnarray}
Now, we use l'Hospital's rule, re-substitute $b$ and $q$, make use
of $b<\infty$ and get \eq{theoEs1}. Note that for any bound on the
amplitude of the signal set $\cal B$, the parameter $q$ is finite.
Even without bound, $q$ will remain finite for a well-defined
minimization problem. The parameter $b$ behaves in a more
complicated manner. It can be both zero, finite, and infinite as
$\beta\to\infty$ depending on the particular R-transform and the
signal sets ${\cal B}_s$. For $\beta\not\in(0,\infty)$, the
saddle-point limits have to be re-considered.


\section{Particular Results}
\label{PR}

The general result leaves us with two components to specify: 1)
The statistics of the random matrix entering the energy per symbol
via its R-transform. 2) The relaxed signal alphabets ${\cal
B}_{s}\,\forall s\in{\cal S}$. While the relaxed alphabets
characterize a particular method of precoding, the random matrix
statistics depends on the wireless communication system. In the
following, we will consider the following choice for the
statistics of the random matrix.

Consider a vector-valued communication system. Let the received
vector be given as
\begin{equation}
\matr r = \matr {Ht} + \matr n
\end{equation}
where $\matr n$ is white Gaussian noise. Let the components of the
transmitted and received vectors be signals sent and received at
different antenna elements, respectively.

We want to ensure that the received signal is (up to additive
noise) identical to the data vector. This design criteria leads us
to choose the precoding matrix
\begin{equation}
\matr T=\matr H^\dagger \left(\matr{HH}^\dagger\right)^{-1}.
\end{equation}
This means that we invert the channel and get $\matr r=\matr
x+\matr n$ if the matrix inverse exists. This allows to keep the
signal processing at the receiver at a minimum. This is
advantageous if the receiver shall be a low-cost or
battery-powered device.

To model the statistics of the entries of $\matr H$ is a
non-trivial task and a topic of ongoing research, see e.g.\
\cite{debbah:05a} and references therein. For sake of convenience,
we choose in this first order approach that the entries of the
channel matrix $\matr H$ are independent and identically
distributed complex Gaussian random variables with zero mean and
variance $1/N$. For that case, we find in the appendix that
\begin{eqnarray}
\label{Rinverse}
R(w)&=&\frac{1-\alpha-\sqrt{(1-\alpha)^2-4\alpha w}}{2\alpha w}\\
\label{Rinverseprime}
R^\prime(w)&=&\frac{\left(1-\alpha-\sqrt{(1-\alpha)^2-4\alpha
w}\right)^2}{4\alpha w^2 \sqrt{(1-\alpha)^2-4\alpha w}}.
\end{eqnarray}
It also turns out helpful to recognize that
\begin{equation}
\frac{R^2(w)}{R^\prime(w)} = \frac{\sqrt{(1-\alpha)^2-4\alpha
w}}\alpha.
\end{equation}

In the following, we compare the performances of several
constructions of the redundant signal re-presentations for the
channel model specified above.

\subsection{1-Dimensional Lattice}

Consider binary one-dimensional modulation. Let
\begin{eqnarray}
{\cal S} &=& \{0,1\}\\
{\cal B}_1 = -{\cal B}_{0} &\subset& \RR.
\end{eqnarray}
This is the mathematical description of binary phase-shift keying
on the real line in context of vector precoding.

Using the specifications above, we find in the limit
$\beta\to\infty$
\begin{eqnarray}
\!\!q \!\!&=&\!\!  \int\limits_{\RR} \left|\argmin_{x\in {\cal B}_{1}}\left|z\sqrt{\frac{qR^\prime(-b)}{2R^2(-b)}}-x\right|\right|^2 \frac{{\rm e}^{-\frac{z^2}2}{\rm d}z}{\sqrt{2\pi}}\\
\!\!b \!\!&=&\!\!\int\limits_{\RR} \argmin_{x\in {\cal
B}_{1}}\left|z\sqrt{\frac{qR^\prime(-b)}{2R^2(-b)}}-x\right|\frac{z\,{\rm
e}^{-\frac{z^2}2}{\rm d}z}{\sqrt{4\pi qR^\prime(-b)}}.
\end{eqnarray}

Moreover, let without loss of generality
$-\infty=c_0<c_1<\cdots<c_L<c_{L+1}=+\infty$ and
\begin{equation}
{\cal B}_1 = \{c_1,c_2,\dots,c_L\}
\end{equation}
This case describes Tomlinson-Harashima precoding
\cite{tomlinson:71,harashima:72} with optimization over $L$
different representations for each information bit. An example of
such a respresentation for integer lattice points is shown in
Fig.~\ref{1dlattice}. The boundary points of the Voronoi regions
are
\begin{equation}
v_i = \frac{c_i+c_{i-1}}2 \qquad
\end{equation}
and the fixed-point equations for $q$ and $b$ become
\begin{eqnarray}
q &=& \frac1{\sqrt{2\pi}}\sum\limits_{i=1}^{L}{\int\limits_{\frac{\sqrt2 R(-b)v_i}{\sqrt{qR^\prime(-b)}}}^{\frac{\sqrt2 R(-b)v_{i+1}}{\sqrt{qR^\prime(-b)}}} c_{i}^2\,{\rm e}^{-\frac{z^2}2}{\rm d}z}\\
&=& c_1^2+\sum\limits_{i=2}^L \left(c_i^2-c_{i-1}^2\right){\rm Q}\left(\frac{R(-b)(c_i+c_{i-1})}{\sqrt{2qR^\prime(-b)}}\right)\\
b &=& \frac{\sum\limits_{i=2}^L
\left(c_i-c_{i-1}\right)\exp\left(-{\frac{R^2(-b)(c_i+c_{i-1})^2}{4qR^\prime(-b)}}\right)}{\sqrt{4\pi
qR^\prime(-b)}}
\end{eqnarray}
with $\Q=\int_x^\infty \exp(-x^2/2){\rm d}x/\sqrt{2\pi}$ denoting
the Gaussian probability integral.

For the case of no precoding at all, i.e.\ $L=1$, we get
\begin{eqnarray}
b&=& 0\\
q&=& c_1^2\\
E_{\rm s} &=& c_1^2R(0).
\end{eqnarray}

For the case of general $L$, we first restrict to the special case
of a square channel matrix. The rectangular case is addressed
subsequently.

\subsubsection{Square Channel Matrix}

For $\alpha=1$, \eqs{Rinverse}{Rinverseprime} respectively
simplify to
\begin{eqnarray}
\label{rw1}
R(w)&=&\frac1{\sqrt{-w}}\\
R^\prime(w)&=&\frac1{2(-w)^{\frac32}}. \label{rw1prime}
\end{eqnarray}
Thus, we find
\begin{equation}
E_{\rm s} \to\infty \qquad {\rm if} \lim\limits_{\beta\to\infty} b
= 0.
\end{equation}
For positive values of $b$, we get
\begin{eqnarray}
q &=& c_1^2+\sum\limits_{i=2}^L \left(c_i^2-c_{i-1}^2\right){\rm Q}\left(b^{\frac14}q^{-\frac12}(c_i+c_{i-1})\right)\\
b &=& \frac{b^{\frac34}}{\sqrt{2\pi q}} \sum\limits_{i=2}^L
\left(c_i-c_{i-1}\right)\exp\left(-\frac{\sqrt
b(c_i+c_{i-1})^2}{2q}\right).
\end{eqnarray}
and
\begin{equation}
E_{\rm s} = \frac q{2\sqrt b}
\end{equation}
which makes the case distinction with respect to the asymptotic
behavior of $b$ obsolete. Moreover, we can combine the above 3
equations to find
\begin{equation}
\label{fixsqinv} E_{\rm s} = \pi
\left[\frac{c_1^2+\sum\limits_{i=2}^L
\left(c_i^2-c_{i-1}^2\right){\rm
Q}\left(\frac{c_i+c_{i-1}}{\sqrt{2E_{\rm
s}}}\right)}{\sum\limits_{i=2}^L
\left(c_i-c_{i-1}\right)\exp\left(-\frac{(c_i+c_{i-1})^2}{4E_{\rm
s}}\right)}\right]^2.
\end{equation}
Numerical solutions to \eq{fixsqinv} are shown in
Table~\ref{tabsqinv}
\begin{table}[p]
\caption{\label{tabsqinv}Energy per symbol for inverted square
channel.}
\begin{center}
\begin{tabular}{|l||c|c|c|c|c|}
\hline
$L$ & 1 & 2 & 3 & 4 & $\infty$\\
\hline
$E_{\rm s}$ & $\infty$ & 2.6942 & 2.6656 & 2.6655 & 2.6655\\
\hline
$E_{\rm s}$ [dB] & $\infty$ & 4.3043 & 4.2579 & 4.2578 & 4.2578\\
\hline
\end{tabular}
\end{center}
\end{table}
for the equally spaced integer lattice
\begin{equation}
{\cal B}_0 = \{+1,-3,+5,-7,+9,\dots\} \label{intlattice}
\end{equation}
and various numbers of lattice points. Obviously, there is little
improvement when going from two to three lattice points and
negligible improvement for more than 3 lattice points.

\subsubsection{Rectangular Channel Matrix}

For a rectangular channel matrix, the Gramian is only invertible
for $\alpha\le1$. However, the R-transform is well-defined for any
positive aspect ratio. For singular random matrices, the
R-transform reflects the fact that the asymptotic eigenvalue
distribution has some point mass at infinity.

Thus, we find
\begin{eqnarray*}
q &=&  c_1^2+\sum\limits_{i=2}^L \left(c_i^2-c_{i-1}^2\right){\rm Q}\left(\textstyle\frac{\left((1-\alpha)^2+4\alpha b\right)^{\frac14}(c_i+c_{i-1})}{\sqrt{2q\alpha}}\right)\\
b &=& \frac{b\sqrt{\frac\alpha{\pi q} \sqrt{(1-\alpha)^2+4\alpha
b}}}{\alpha-1+\sqrt{(1-\alpha)^2+4\alpha b}}
\sum\limits_{i=2}^L \left(c_i-c_{i-1}\right){\rm
e}^{-{\frac{\sqrt{(1-\alpha)^2+4\alpha
b}(c_i+c_{i-1})^2}{4q\alpha}}}.
\end{eqnarray*}
It is convenient to replace the parameter $b$ by the substitution
\begin{equation}
p= \sqrt{(1-\alpha)^2+4\alpha b}
\end{equation}
which gives
\begin{eqnarray*}
q &=&  c_1^2+\sum\limits_{i=2}^L \left(c_i^2-c_{i-1}^2\right){\rm Q}\left(\sqrt{\frac{p}{2q\alpha}}(c_i+c_{i-1})\right)\\
p &=& 1-\alpha + {\sqrt{\frac{\alpha p}{\pi q} }}
\sum\limits_{i=2}^L
\left(c_i-c_{i-1}\right)\exp\left(-{\frac{p(c_i+c_{i-1})^2}{4q\alpha}}\right)
\end{eqnarray*}
and
\begin{equation}
\label{Es1dimInv} E_{\rm s} = \frac qp.
\end{equation}
Finally, combining the last three equations, we get
\begin{equation}
E_{\rm s} = \frac{  c_1^2+\sum\limits_{i=2}^L
\left(c_i^2-c_{i-1}^2\right){\rm
Q}\left(\frac{c_i+c_{i-1}}{\sqrt{2\alpha E_{\rm
s}}}\right)}{1-\alpha + {\sqrt{\frac{\alpha}{\pi E_{\rm s}} }}
\sum\limits_{i=2}^L
\left(c_i-c_{i-1}\right)\exp\left(-{\frac{(c_i+c_{i-1})^2}{4\alpha
E_{\rm s}}}\right)}
\end{equation}

\begin{figure}
\centerline{\epsfig{file=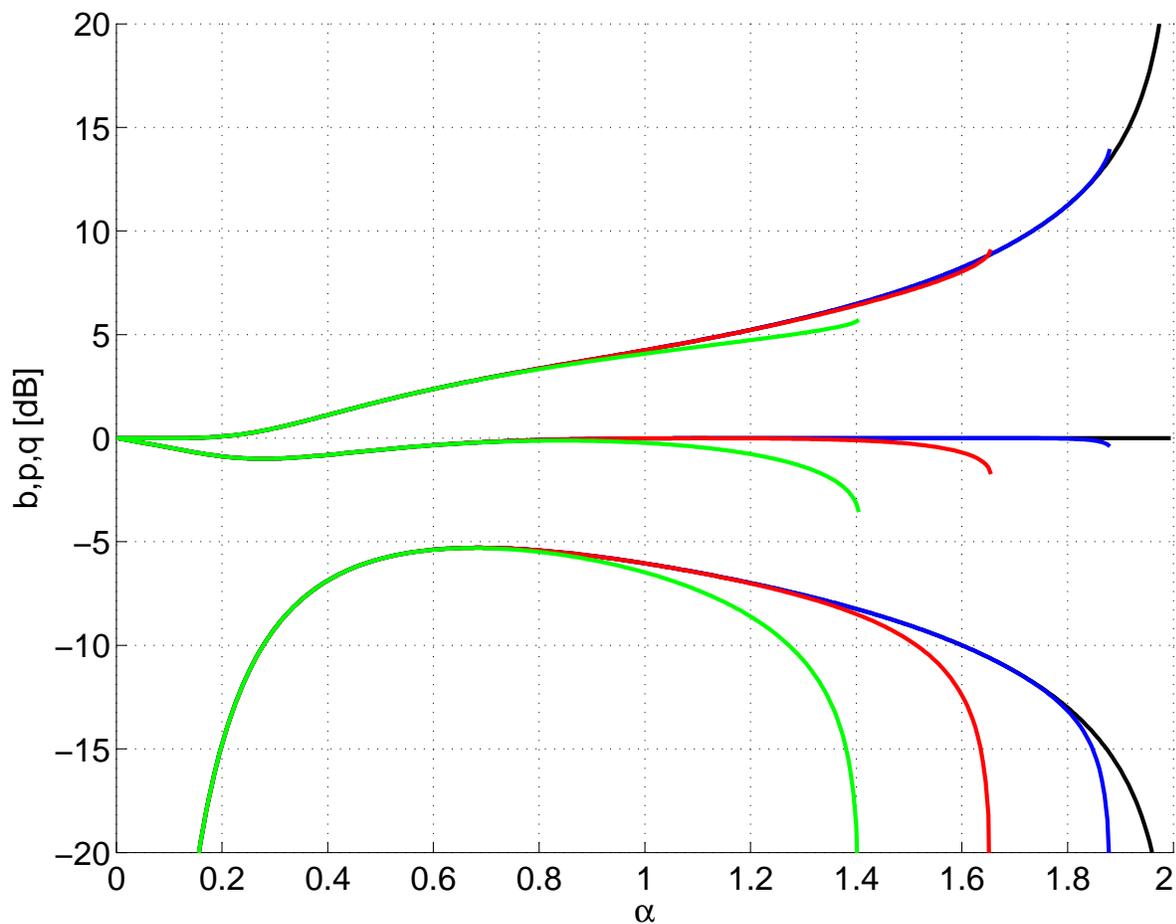,width=\columnwidth}}
\caption{\label{q_and_m_rect} The macroscopic parameters $q$
(upper lines), $b$ (lower lines), and $p$ (medium lines) versus
the load $\alpha$ for $L=2,3,6,100$. shown by green, red, blue,
and black lines, respectively.}
\end{figure}

\begin{figure}
\centerline{\epsfig{file=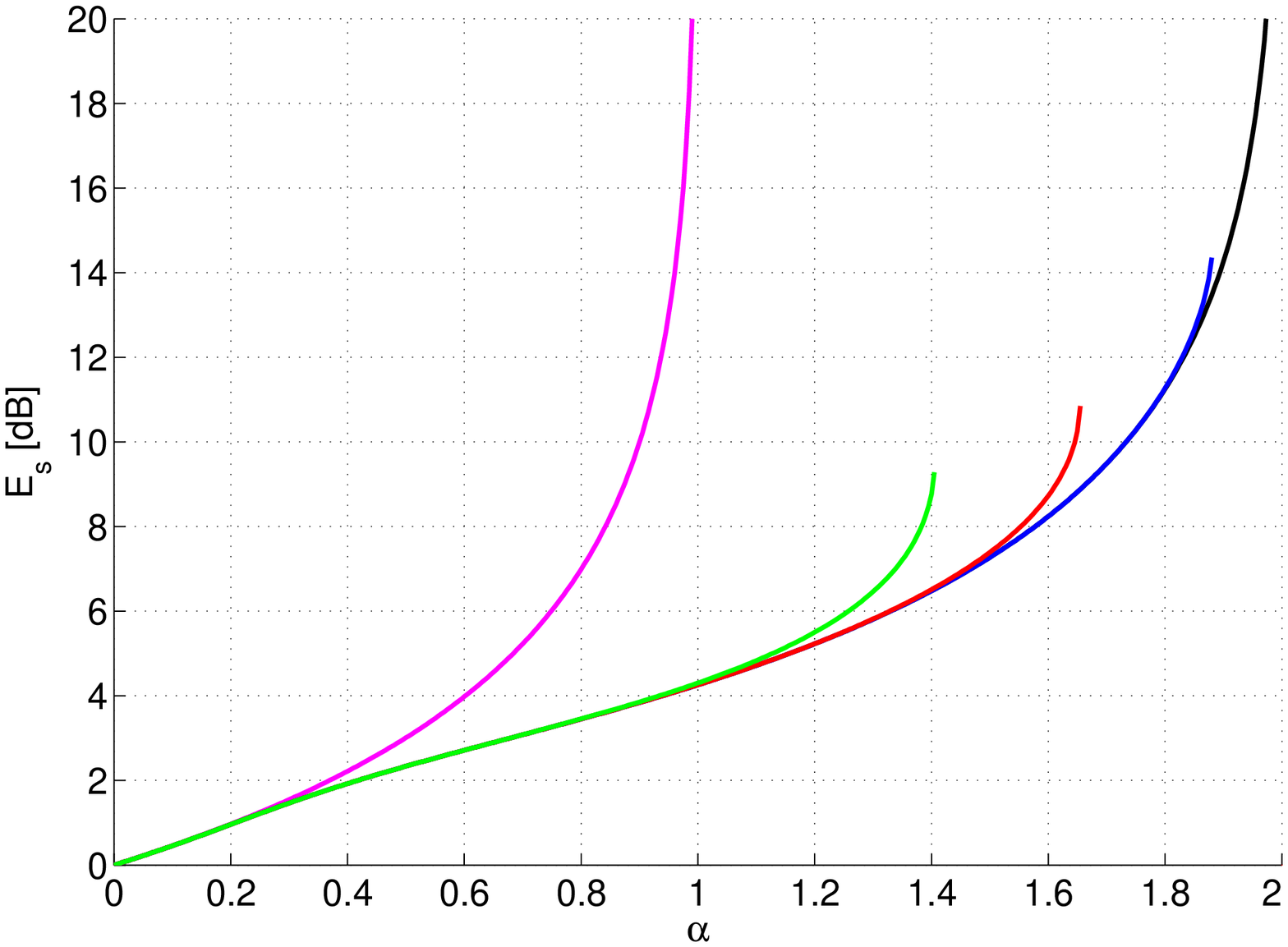,width=\columnwidth}}
\caption{\label{rect} The transmitted energy per symbol versus the
load for  $L=1,2,3,6,100$ shown by the magenta, green, red, blue,
and black lines, respectively.}
\end{figure}

The solutions of these fixed-point equations are shown by the
solid lines in Fig.~\ref{q_and_m_rect}.
Clearly for small load, the parameter $q$ tends to 1, i.e.\ 0 dB,
as in that case, no gain due to precoding is possible and the
symbol with smallest magnitude is preferred. The minimum of the
transmit power is shown by the solid line in Fig.~\ref{rect}.
Note that precoding enables to achieved finite transmitted energy
per symbol even if the channel matrix is singular. This effect has
already been explained for Marchenko-Pastur distributed random
matrices. Unlike the curve without precoding, the curves for $L>1$
do not have poles at the threshold load. Instead, a phase
transition occurs and the energy per symbol jumps discontinuously
from a finite value to infinity. In fact, it can be shown that the
threshold load at which this happens is universal for a large
class of random matrices in that it depends only on the specifics
of the precoding lattice but not on the channel statistics.

\subsection{2-Dimensional Quadrature Lattice}
\begin{figure}[p]
\centerline{\epsfig{file=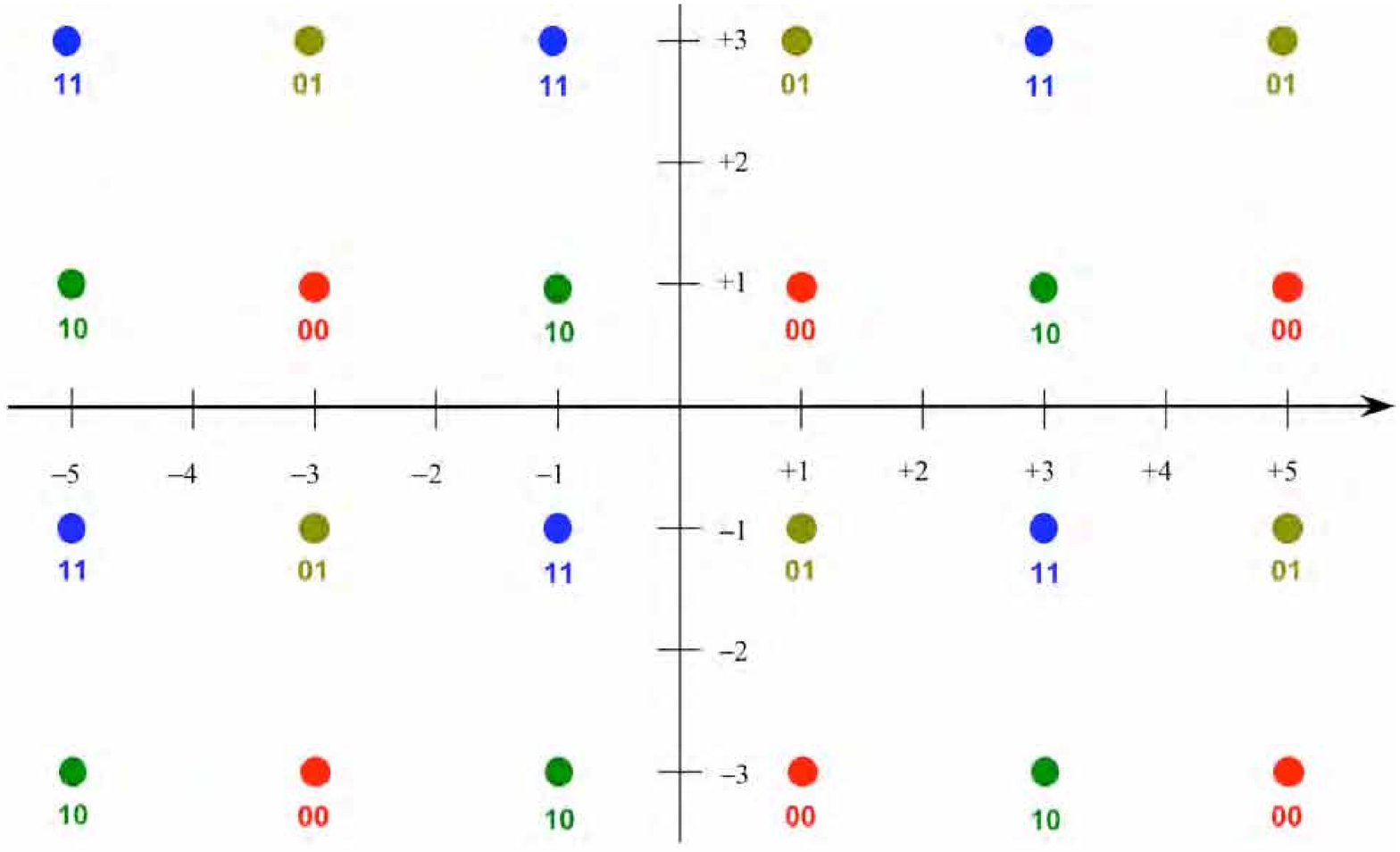,width=\columnwidth}}
\caption{\label{figTH4Q} 4 two-dimensional equally spaced integer
quadrature lattices representing the four quaternary states 00,
01, 10, and 11, respectively.}
\end{figure}

Consider now the case shown in Fig.~\ref{figTH4Q}.
It has the following properties:
\begin{eqnarray}
{\cal S} &=& \{00,01,10,11\}\\
{\cal B}_{1y} &=& - {\cal B}_{0y}^\ast \qquad \forall y\in\{0,1\}\\
{\cal B}_{x1} &=& + {\cal B}_{x0}^\ast \qquad \forall x\in\{0,1\}
\end{eqnarray}
This case extends the one-dimensional precoding of binary
phase-shift keying (BPSK) on the real line to two-dimensional
precoding of quaternary phase-shift keying (QPSK) in the complex
plane such that Gray mapping is applied and we can consider the
pre-coding for QPSK as independent pre-coding of BPSK in both
quadrature components.

The symmetry in both quadrature components implies that
\begin{eqnarray}
q\!\! &=& \!\!\sqrt{\frac2\pi} {\int\limits_{\RR} \!\!\left|\argmin_{x\in \Re\left\{{\cal B}_{1+\j}\right\}}\left|z\sqrt{\frac{qR^\prime(-b)}{2R^2(-b)}}-x\right|\right|^2 {\rm e}^{-\frac{z^2}2}{\rm d}z}\\
b \!\!&=&\!\!\!\!\int\limits_{\RR}\!\! \argmin_{x\in
\Re\left\{{\cal
B}_{1+\j}\right\}}\left|z\sqrt{\frac{qR^\prime(-b)}{2R^2(-b)}}-x\right|\frac{z\,{\rm
e}^{-\frac{z^2}2}{\rm d}z}{\sqrt{\pi qR^\prime(-b)}}.
\end{eqnarray}
Compared to the one-dimensional case, the only difference is that
the right hand sides of the two fixed point equations are
multiplied by a factor of 2 which stems from adding the
contributions of both quadrature components. In order to allow for
a fair comparison with 1-dimensional modulation, we shall consider
the energy per bit
\begin{equation}
E_{\rm b}=\frac{E_{\rm s}}{\log_2\left|{\cal S}\right|}
\end{equation}
to be the performance measure of choice.

\begin{figure}[p]
\centerline{\epsfig{file=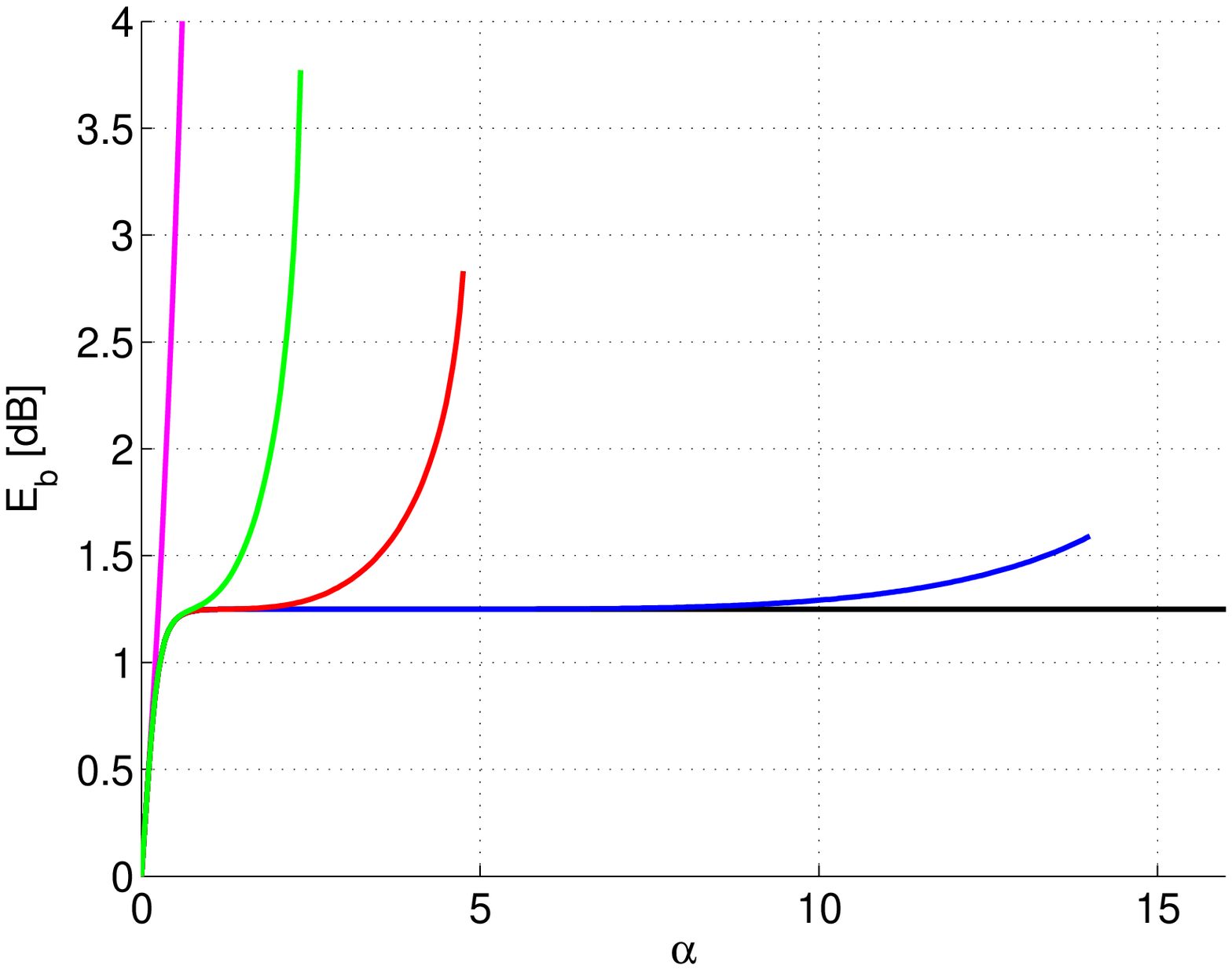,width=\columnwidth}}
\caption{\label{GrayInv} Transmitted energy per bit versus the
load for channel inversion and pre-coding for Gray-mapped QPSK
with $L=1,2,3,6,100$ shown by the magenta, green, blue, and black
lines respectively.}
\end{figure}

Due to de-coupling between quadrature components, we find that
\eq{Es1dimInv} remains valid and $p$ and $q$ are given by
\begin{eqnarray}
q\!\! &=&\!\!  2c_1^2+2\sum\limits_{i=2}^L \left(c_i^2-c_{i-1}^2\right){\rm Q}\left(\sqrt{\frac{p}{2q\alpha}}(c_i+c_{i-1})\right)\\
p\!\! &=&\!\! 1-\alpha + {\sqrt{\frac{4\alpha p}{\pi q} }}
\sum\limits_{i=2}^L \left(c_i-c_{i-1}\right){\rm
e}^{-{\frac{p(c_i+c_{i-1})^2}{4q\alpha}}}.
\end{eqnarray}
The solutions to these fixed point equations are shown in
Fig.~\ref{GrayInv}.
Remarkably, the energy per bit remains as small as $E_{\rm
b}=\frac43$ for any load if $L$ grows large.

\subsection{2-Dimensional Checkerboard Lattice}

\begin{figure}[p]
\centerline{\epsfig{file=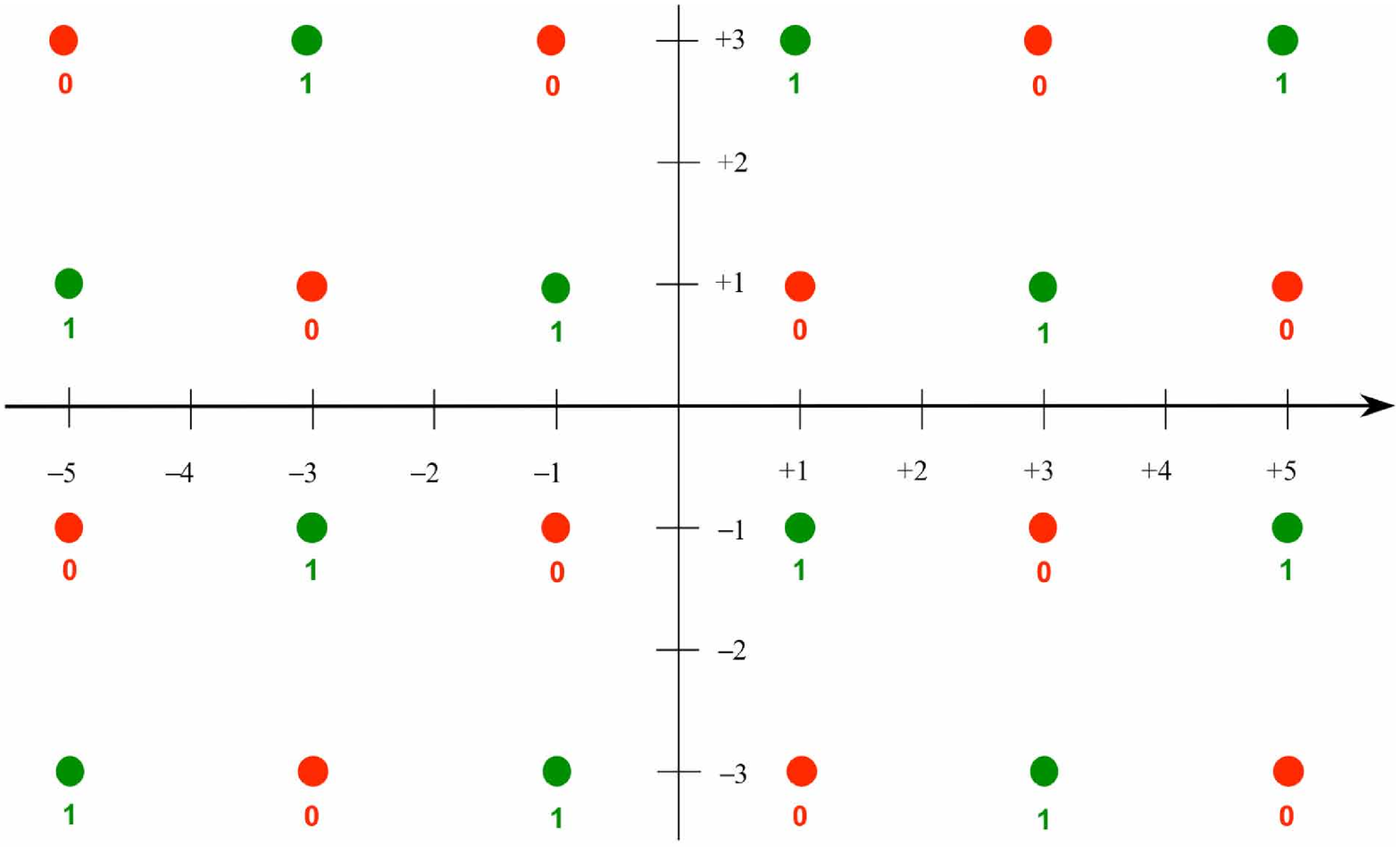,width=\columnwidth}}
\caption{\label{figTH4CB} 2 two-dimensional equally spaced integer
quadrature lattices representing the two binary states 0 and 1,
respectively.}
\end{figure}

Consider now the case shown in Fig.~\ref{figTH4CB}.
This case extends the one-dimensional pre-coding of BPSK on the
real line to two-dimensional pre-coding of BPSK in the complex
plane. Among others, it has the following properties:
\begin{eqnarray}
{\cal S} &=& \{0,1\}\\
{\cal B}_1 = \j {\cal B}_{0} &\subset& \CC.
\end{eqnarray}

This mapping is like a checkerboard where the sets ${\cal B}_1$
and ${\cal B}_{0}$ correspond to the black and white fields,
respectively. For this mapping, the boarderlines of the Voronoi
regions are not parallel to the real and imaginary axes but
intersect these by an angle of $45^{\rm o}$.

Considering an unconstrained lattice, i.e.\ infinitly many lattice
points, we can rotate the lattice by $45^{\rm o}$ degrees without
loss of generality due to the rotational invariance of the complex
Gaussian integral kernel in the fixed-point equations for $b$ and
$q$. After rotation we find the same lattice as in the
two-dimensional quadrature precoding except for a lattice scaling
by a factor of $1/\sqrt2$. Thus, the energy per symbol will be
half the energy per symbol of quadrature precoding and the energy
per bit will be identical.

\subsection{2-Dimensional Semi-Discrete Lattice}
\begin{figure}[pp!tb]
\centerline{\epsfig{file=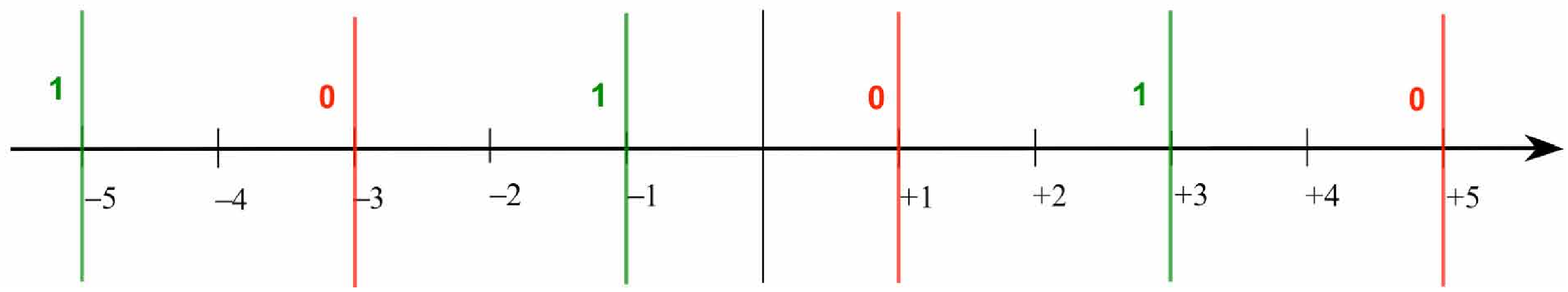,width=\columnwidth}}
\caption{\label{figTH4SD} 2 two-dimensional equally spaced
semi-discrete lattices representing the two binary states 0 and 1,
respectively.}
\end{figure}

Consider now the case shown in Fig.~\ref{figTH4SD}.
This mapping is identical to the 1-dimensional lattice except for
the fact that the imaginary parts of the symbols in ${\cal B}_{x}$
are left arbitrary.

This mapping has the following properties
\begin{eqnarray}
{\cal S} &=& \{0,1\}\\
{\cal B}_1 = - {\cal B}_{0} &\subset& \CC
\end{eqnarray}
which lead to
\begin{eqnarray}
q &=&\frac{qR^\prime(-b)}{2R^2(-b)}
  +  \int\limits_{\RR} \left|\argmin_{x\in \Re\left\{{\cal B}_{1}\right\}}\left|z\sqrt{\frac{qR^\prime(-b)}{2R^2(-b)}}-x\right|\right|^2
\frac{{\rm e}^{-\frac{z^2}2}{\rm d}z}{\sqrt{2\pi}}\\
b &=&\frac1{2R(-b)}
 +\int\limits_{\RR} \argmin_{x\in \Re\left\{{\cal B}_{1}\right\}}\left|z\sqrt{\frac{qR^\prime(-b)}{2R(-b)^2}}-x\right|\frac{z\,{\rm e}^{-\frac{z^2}2}{\rm d}z} {\sqrt{4\pi qR^\prime(-b)}}.
\end{eqnarray}
For channel inversion, we have
\begin{equation}
\frac{R^\prime(-b)}{R^2(-b)} = \frac\alpha p.
\end{equation}
This enables us to easily solve the fixed point equations.

\begin{figure}[tb]
\centerline{\epsfig{file=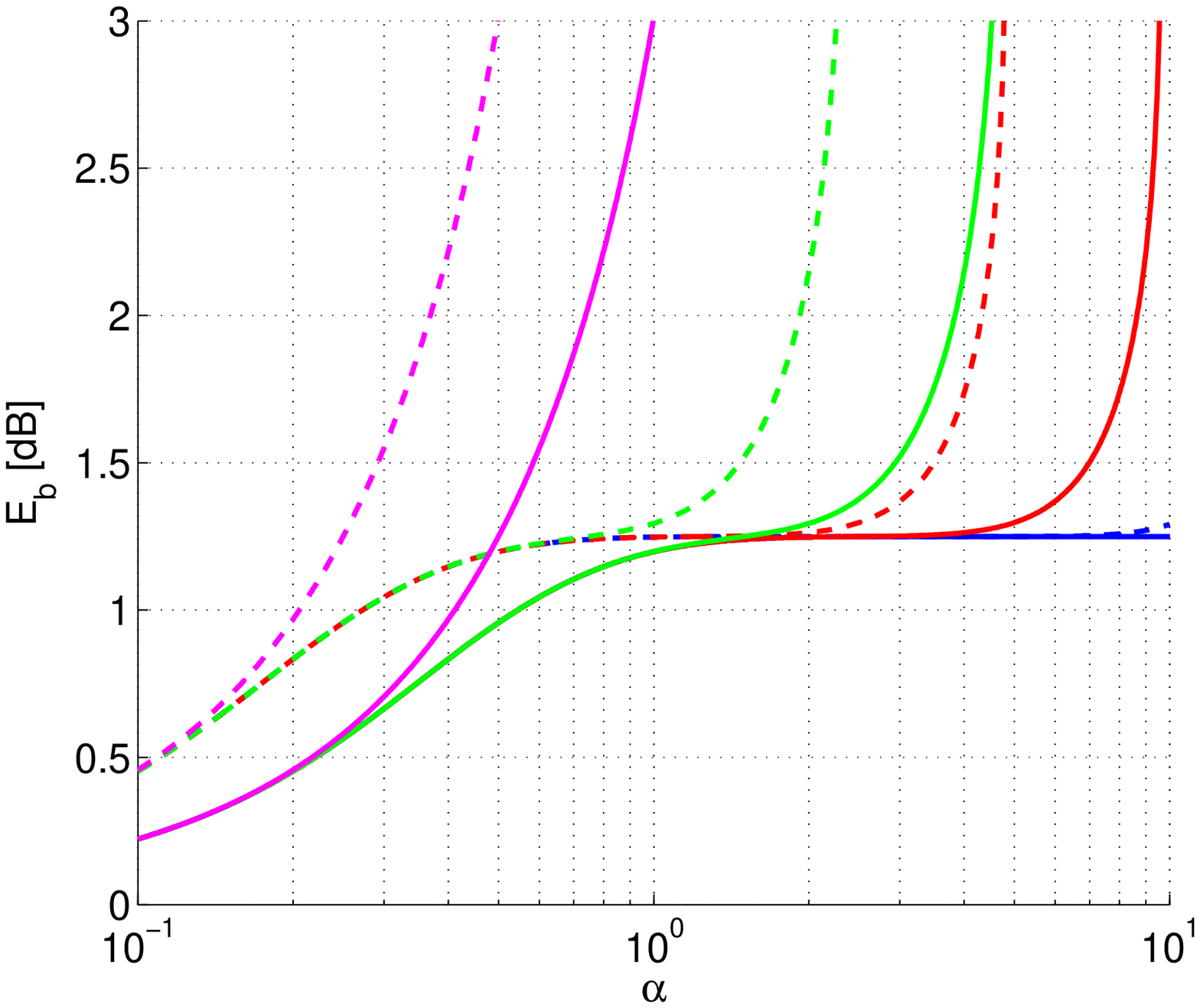,width=\columnwidth}}
\caption{\label{invMPcomp} Energy per bit versus load for
precoding with complex quadrature lattice (dashed lines) and
semi-discrete lattice (solid lines) for $L=1,2,3,6$ shown by the
magenta, green, red, and blue lines, respectively.}
\end{figure}

Fig.~\ref{invMPcomp}
compares the complex semi-discrete lattice with the complex
quadrature lattice in terms of energy per bit. Precoding with
semi-discrete lattices achieves a remarkable gain which comes at
the expense of reduced data rate. It is particularly worth to
remark that the semi-discrete lattice with $L=1$ outperforms all
quadrature lattices for loads up to $\alpha\approx 0.479$. Note
that for $L=1$, the sets ${\cal B}_s$ are convex. Thus, the
quadratic programming problem is convex since $\matr J$ is
positive semidefinite and it can be solved with a polynomial time
algorithm \cite{boyd:04}. For large loads and large lattice size,
the energy per bit approaches $E_{\rm b} =\frac43$.

\section{Conclusions}
\label{C}

We have found that vector pre-coding can significantly reduce the
required transmitted power. In fact, with appropriate pre-coding,
the transmitted power stays always finite. Moreover, we found
strong advantages of complex-valued pre-coding over real-valued
pre-coding and a trade-off between data rate and required transmit
power.

We are aware of the fact that replica symmetry might not hold.
Therefore, we have started investigating first order replica
symmetry breaking (1RSB). The quantitative analysis is not
finished yet, but qualitatively, the results remain unchanged for
1RSB.

\section*{Acknowledgments}

This research was supported by the Research Council of Norway, the
National Science Foundation, DARPA, and the European Commission
under grants 171133/V30, CCF-0644344, W911NF-07-1-0028, and
MIRG-CT-2005-030833, resp. It was initiated while R. M\"uller and
D. Guo were visiting the Institute for Mathematical Sciences at
the National University of Singapore in 2006.

\section*{Appendix}

Let $\prob Xx$ be an arbitrary pdf such that both the Stieltjes
transform defined in \eq{defST} and
\begin{equation}
m_{X^{-1}}(s)=\int\frac{{\rm d}\Prob Xx}{\frac 1x-s}
\end{equation}
exist for some complex $s$ with $\Im(s)>0$. It can easily be
checked that
\begin{equation}
m_{X^{-1}}\left(\frac1s\right) = -s\left(1+s{m_X(s)}\right).
\end{equation}
Let $s=m_X^{-1}(-w)$. Then, we find
\begin{equation}
m_{X^{-1}}\left(\frac1{m_X^{-1}(-w)}\right) =
-m_X^{-1}(-w)\left(1-wm_X^{-1}(-w)\right).
\end{equation}
and
\begin{equation}
\frac1{m_X^{-1}(-w)} =
m_{X^{-1}}^{-1}\left(-m_X^{-1}(-w)\left(1-wm_X^{-1}(-w)\right)\right).
\end{equation}
With Definition 1, we find
\begin{eqnarray}
\frac{1}{R_X(w)+\frac1w}&=&
R_{X^{-1}}\left(-wR_X(w)\left(R_X(w)+\frac1w\right)\right)
- \frac1{wR_X(w)\left(R_X(w)+\frac1w\right)}
\end{eqnarray}
and
\begin{equation}
\label{thelemma} \frac1{R_X(w)} =
R_{X^{-1}}\left(-R_X(w)\left(1+wR_X(w)\right)\right).
\end{equation}

It is well known \cite{voiculescu:92,tulino:04} that for an
$N\times \alpha N$ random matrix $\matr H$ with i.i.d.\ entries of
variance $(\alpha N)^{-1}$, the R-transform of the limiting
spectral measure $\Prob{\matr {H}^{\dagger}\matr H}x$ is given by
\begin{equation}
R_{\matr {H}^{\dagger}\matr H}(w)=\frac 1{1-\alpha w}.
\end{equation}
Letting $X^{-1} = \matr{H}^{\dagger}\matr H$, we find
\begin{equation}
{R_{(\matr {H}^{\dagger}\matr H)^{-1}}(w)} = {1+\alpha R_{(\matr
{H}^{\dagger}\matr H)^{-1}}(w)\left(1+wR_{(\matr
{H}^{\dagger}\matr H)^{-1}}(w)\right)} \label{eq111}
\end{equation}
with \eq{thelemma}. Solving \eq{eq111} for the R-transform implies
\eq{Rinverse}. Note that for $\alpha\ge1$, the mean of the
spectral measure is diverging. Thus, the R-transform must have a
pole at $w=0$ which excludes the other solution of \eq{eq111}.


\end{document}